\documentclass[a4paper,aps,pra,floats]{revtex4}
\usepackage{overpic}
\usepackage{graphicx, latexsym, verbatim}
\usepackage{graphics}
\usepackage{amssymb, amsmath}
\usepackage[ansinew]{inputenc}
\usepackage{color}
\usepackage{multirow}
\definecolor{darkblue}{rgb}{0,0,0.55}
\definecolor{comfrey}{rgb}{0.85,0.85,0.85}
\usepackage{rotating}
\newcommand{\ud}{\mathrm{d}}
\newcommand{\mr}{\mathbf{r}}

\newcommand{\mE}{\mathbf{E}}

\newcommand{\mft}{\tilde{\mathbf{f}}}

\newcommand{\mgt}{\tilde{\mathbf{g}}}
\newcommand{\tlo}{\tilde{\omega}}

\newcommand{\me}{\mathbf{e}}

\definecolor{myBlue}{rgb}{0.0430,0.5156,0.7773}

\newcommand{\red}[1]{{\color{black}#1}}
\newcommand{\green}[1]{{\color{green} }}
\newcommand{\ptk}[1]{{\color{black}#1}}
\newcommand{\ptkTwo}[1]{{\color{black}#1}}

\begin{document}

% reduce the space around align equations
%\abovedisplayskip=7pt
%\abovedisplayshortskip=0pt
%\belowdisplayskip=7pt
%\belowdisplayshortskip=7pt

\bibliographystyle{thesis_bibliographystyle}
% Bibliography styles that can be used instead of prsty are abbrv, alpha, plain and unsrt

%Title of paper
\title{Reply to ``Comment on ``Normalization of quasinormal modes in leaky optical
    cavities and plasmonic resonators'' ''
    by E. A. Muljarov and W. Langbein}
\author{Philip Tr\o st Kristensen}
\affiliation{Institut f\"ur Physik, Humboldt Universit\"at zu Berlin, 12489 Berlin, Germany}
\author{Rong-Chun Ge}
\author{Stephen Hughes}
%\affiliation{$^\dagger$Division of Physics and Applied Physics, Nanyang Technological University, 637371, Singapore.}
\affiliation{Department of Physics, Queen's University, Kingston, Ontario, Canada, K7L 3N6}

\date{\today}

%\begin{center}
%{\bf\large Re: APK1022}\\
%\end{center}

\begin{abstract}
We refute all claims of the ``Comment on ``Normalization of quasinormal modes in leaky optical cavities and plasmonic resonators''\,'' by E. A. Muljarov and W. Langbein (arXiv:1602.07278v1). Based entirely on information already contained in our original article (P. T. Kristensen, R.-C. Ge and S. Hughes, Physical Review A \textbf{92}, 053810 (2015)), we dismiss every point of criticism as being completely unjustified and point out how important parts of our argumentation appear to have been overlooked by the Comment Authors. %and illustrate how our original results and discussions appear to have been either overlooked or misunderstood by the Comment authors. %\green{We explain why our results are correct and why this Comment, sadly, amounts to nothing more than petty nitpicking.} 
In addition, we provide additional calculations showing directly the connection between the normalizations by Sauvan \emph{et al.} and Muljarov \emph{et al.}, which were not included in our original article.
\end{abstract}

\maketitle

In a recent article~\cite{1}, we have put forward the point of view that three seemingly different normalizations for so-called quasinormal modes (QNMs) can be understood as arising from different procedures for regularization of an inherently ill-behaved integral. We have done this from the general point of view \ptkTwo{that the three normalizations are different formulations of the same quantity (the norm)}, providing calculations that show how one can derive one formulation from the other. Moreover, we have provided explicit calculation examples for three different material systems of current interest in the literature. In a recent arXiv submission entitled ``Comment on ``Normalization of quasinormal modes in leaky optical cavities and plasmonic resonators''\,''~\cite{arXiv:1602.07278v1}, the Comment authors raise criticism of our work which, in view of the %upon comparison with 
content of Ref.~\cite{1}, we find is completely unjustified. %can be completely dismissed. %easily and completely dismissed. 

Before addressing the criticism in the Comment, we briefly outline the motivation for Ref.~\cite{1} and its context with other publications concerning the normalization of QNMs. The work in Ref.~\cite{1} was partly motivated by an earlier arXiv submission from Muljarov, Doost and Langbein~\cite{2a}, which was recently updated to Version 3, now with only  Muljarov and Langbein~\cite{2}. % \ptk{these works in turn build on earlier results on perturbation theory in open systems~\cite{EPL} based on a norm for QNMs which is arguably the most elegant of the three formulations in question.}
Building on earlier results on perturbation theory in open systems~\cite{EPL}, Ref.~\cite{2a} % , and a norm for QNMs which is arguably the most elegant of the three formulations in question, Muljarov (Doost) and Langbein derived results that
discusses how one can use the theory of QNMs to define the correct mode volume for use %a so-called generalized effective mode volume for use 
in Purcell factor calculations in open optical systems. Exactly the same conclusion %The same conclusion, 
%and the notion of a generalized effective mode volume, %A similar conclusion 
was previously %had %result had 
%previously been 
presented in Ref.~\cite{OL} in terms of a generalized effective mode volume and based on a %
%In essence, this is %\green{exactly} 
%the same conclusion that was previously presented in Ref.~\cite{OL} using a %result that was previously published by two of us~\cite{OL} (Kristensen and Hughes, with Van Vlack) using a 
normalization for QNMs due to Lai and co-workers~\cite{Lai}. Parts of Ref.~\cite{2a}, therefore, is devoted to the claim that the normalization used in Ref.~\cite{OL} is incorrect owing to a difficulty in the practical evaluation of the norm, which is evident for QNMs in cavities with (very) low $Q$-values. \ptk{The work in Ref.~\cite{1} was also} partly inspired by a recent %independent
publication from Sauvan and co-workers~\cite{Sauvan}, who derived results that (for dispersionless materials) appear similar to those in Ref.~\cite{OL} but who did not discuss the connection to other normalization procedures. The link between the normalizations in Refs.~\cite{Lai} and~\cite{Sauvan} %of Lai~\cite{Lai} and Sauvan~\cite{Sauvan} 
was discussed in Ref.~\cite{Ge}, and with %. % (which also included the case of dispersive materials). 
%With 
the submission to the arXiv server of Ref.~\cite{2a}, %therefore, 
it seemed likely that %worth the effort to investigate if
a similar connection could be found for the normalization used in this work also. %is normalization also.
After having found such a connection, we presented the relationship between all three normalizations %in Ref. 
from the general point of view that they are complementary~\cite{1}. At the same time, we provided explicit examples to show that the normalization in Ref.~\cite{OL} is in fact correct and indeed leads to the same result as the other two formulations. %In particular, we included explicit calculations to 
These calculation examples specifically addressed points of criticism in Ref.~\cite{2a} which are practically identical to what are now put forward in Ref.~\cite{arXiv:1602.07278v1} once again. Hence, it is with some reluctance that we engage in this Reply, since all arguments have already been presented. Yet we shall do so nonetheless, and we refer all interested readers to Ref.~\cite{1} for details. 

\red{In the Appendix, we present calculations, which unfortunately we did not manage to include in Ref.~\cite{1}, showing directly the connection between the normalizations in Refs.~\cite{EPL} and~\cite{Sauvan}.}
\newline
%in acceptance of the fact that ...
%after having learned of 
%.. may take the criticism at face value without comparing 
%not everyone has the time to reflect on the criticism in the Comment and may therefore take 

In Ref.~\cite{arXiv:1602.07278v1}, %``Comment on ``Normalization of quasinormal modes in leaky optical cavities and plasmonic resonators''\,'', 
the Comment authors claim:
\begin{itemize}
\item[(a)] that the conclusion in Ref.~\cite{1}, that the three different methods for normalization of QNMs all provide the same result, is incorrect,
\item[(b)] %Moreover, they claim (b) 
that the normalization in Ref.~\cite{OL} is divergent for any optical mode having a finite $Q$ value, and 
\item[(c)] that the Silver-M{\"u}ller (SM) radiation condition is not fulfilled for QNMs.
\end{itemize}
\newpage
\textbf{Concerning (a)}
\newline
%In Ref.~\cite{1}, the point of view was put forward, that three seemingly different normalization formulas are complementary and provide the same results. The Comment authors claim that this is incorrect because of points (b) and (c).
 
In Ref.~\cite{1}, we provided general calculations showing the connection between the three normalizations for general QNMs. In particular, we discussed %at length 
how one can always regularize the normalization integral of Lai and co-workers by use of a complex coordinate transformation. With such an approach, and for general resonators, albeit only in spherical integration domains, we showed how one can rewrite the normalization integral of Lai and co-workers in the exact form favored by the Comment authors. In addition to showing the connection between the normalizations for general resonators, we provided three example calculations, which we consider to be of contemporary interest. In the first example, we considered the exact same QNMs that were previously investigated by the Comment authors in Ref.~\cite{2a}, and we calculated the norm using the three different formulations. The results of those calculations are listed in Eqs. (27), (28) and (30) of Ref.~\cite{1}, and they are identical. The Comment authors appear to have %completely 
overlooked these important parts of our argumentation. For any questioning of the results in Ref.~\cite{1} to carry real weight, however, one would expect an indication of a flaw in the analytical calculations showing the connection between the normalizations for general resonators, or at least % and %or %at % showing the connection between the normalization integrals for general resonators, or at 
%least 
an explanation for the exact equivalence of the results in the example calculations. %of the results in Eqs. (27), (28) and (30) of Ref.~\cite{1}.
In view of our results, and the lack of any such explanation, we find the claim of the Comment authors, that the three normalization methods do not provide the same result, to be completely unjustified.\newline

%our calculations showing equivalence for general resonators as well as the explicit calculation examples, 
%yet for their arguments to carry any meaningful weight, one would expect them % at least %that they would at least provide 
%to be able to point out a flaw in our calculations showing the connection between the normalization integrals for general resonators, or at least an explanation for the exact equivalence of the results in Eqs. (27), (28) and (30) of Ref.~\cite{1}, % in the explicit calculation example, 
%if indeed the conclusion in Ref.~\cite{1}, that all three formulations lead to the same result, is incorrect.\newline
%In view of the general result in Ref.~\cite{1}, which is valid for arbitrary resonators, as well as the explicit calculation examples, we find that the claims of the Comment authors, that the methods do not give the same result, are completely unjustified.\newline

\textbf{Concerning (b)}
\newline

The Comment authors first argue that
%\vspace{0.3cm}
\begin{itemize}
\item[]``... the LK normalization [...] mathematically does not exist,''
\end{itemize} 
essentially pointing out the same technical details as in Ref.~\cite{2a}, which were fully acknowledged and discussed %at length 
in Ref.~\cite{1}, where it was shown how one can in principle always regularize the so-called LK normalization by a complex coordinate transformation, although this relies on the expansion of the QNMs into spherical wave functions and therefore is mostly of theoretical interest unless the QNMs are known analytically. The Comment authors appear to acknowledge that such a regularization is always possible, since they (later) write:
%\vspace{0.3cm}
\begin{itemize}
\item[]``... For this [complex coordinate transformation] to be used, the fields of the [QNMs] have to be known analytically. This regularization is thus not suited for numerically determined [QNMs].''
\end{itemize} 
This statement contradicts the earlier statement that ``... the LK normalization [...] mathematically does not exist''. %, and thus the Comment appears to be internally inconsistent on this point. 
Moreover, we note that the criticism at this point is relaxed somewhat from being a fundamental question of whether or not the normalization mathematically ``exists'', into a question of whether or not the normalization is ``suited for numerically determined [QNMs]''.

To reiterate: As discussed %at length 
in Ref.~\cite{1}, and contrary to the claim by the Comment authors in the above quote, one can in principle always regularize the so-called LK normalization by a complex coordinate transformation (even for numerically determined QNMs by projection onto the spherical wave functions). Therefore, there should be no doubt that the so-called LK normalization ``exists'', and for QNMs that are known analytically, one can easily apply the regularization also in practice. Judging from the above quote, the Comment authors appear to (implicitly) acknowledge this, yet they devote three figures to show that the integral is not well behaved if the radius of the calculation domain is varied along the real axis (as fully acknowledged and discussed at length in Ref.~\cite{1}). Moreover, they show this for QNMs that are in fact known analytically, and for which the proposed regularization is directly and easily applicable. This appears to be a major misunderstanding of our entire discussion in Ref.~\cite{1}.\newline

Later in the Comment's text appears a new argument, namely:
%\vspace{0.3cm}
\begin{itemize}
\item[]``We emphasize that this ``regularized'' LK normalization is a different quantity compared to the divergent LK normalization defined by Eqs. (1-2) what was actually used in [Ref.~\cite{OL}] and numerous follow-up publications of the same group...''
\end{itemize} 
It appears then, at this point in the Comment, that the problem with the so-called LK normalization is no longer the fact that it ``does not exist'', but rather that the regularized normalization is different from what was actually used in Ref.~\cite{OL} and other later publications. The Comment, therefore, appears now to cover the general use of the so-called LK normalization and in particular its use in Ref.~\cite{OL}. When viewed in the context of the claims that this %the so-called LK 
normalization is ``not suited for numerically determined [QNMs]'', it naturally leads to the question of whether the calculated results in Ref.~\cite{OL} and later publications can even be trusted. %Such a suspicion is evident  most clearly in the introductory paragraph, where it is written directly
%\begin{itemize}
%\item[]``a regularized variant of the LK normalization, put forward in [Ref.~\cite{1}], is not suited for numerically determined [QNMs]''
%\end{itemize} 
%Such a suspicion is clearly evident in the above statement that the so-called LK normalization is not ``suited for numerically determined [QNMs]'', which is also w
%so that also the practical usefulness of the so-called LK normalization is now questioned. 
Also in this case, the Comment authors appear to have overlooked the results presented in Ref.~\cite{1}, which in fact included a lengthy discussion about the calculation of the norm for the cavity in Ref.~\cite{OL} with the lowest Q-value (Q=16). From simply reading off the graph (as we did in preparing Ref.~\cite{OL}), we can determine the norm with a relative error better than 0.0001 (one part in ten thousand relative error). By slightly more elaborate methods (as detailed in Ref.~\cite{1}), we can reduce the estimated relative error to less than 0.000002 (two parts in a million relative error). Comparing directly to the normalization favored by the Comment authors, we get the same number to 6 digits accuracy, cf. Ref.~\cite{1}. In view of these results, we consider the claims by the Comment authors, that the two methods do not give the same result, or that the so-called LK normalization ``is not suitable for numerically determined [QNMs]'', to be completely unjustified.\newline

\textbf{Concerning (c)}
\newline

%will require some explanation of this extraordinary agreement in order to carry any weight.
It was argued in Ref.~\cite{1} that the QNMs can be defined as the solutions to the wave equations that fulfill a certain variant of the SM radiation condition. The Comment authors claim that the QNMs do not fulfill this condition because the wavevector is not real. Clearly, if the premise is that the wavevector in the SM condition is real, then the SM condition cannot be fulfilled by QNMs. The same type of argument can be put forward for the wave equation itself. Indeed, one could define the wave equation with the restriction that the wavevector is real, and thereby argue that the QNMs do not fulfill the wave equation. From the context in Ref.~\cite{1} it is clear, however, that no such restriction was made when it was argued that the QNMs should fulfill the SM condition. Moreover, in Ref.~\cite{1} the SM condition was deliberately written in a form which is slightly different from the one presented by the Comment authors; a form for which the arguments put forward in the Comment do not apply, and a form for which the SM condition is in fact fulfilled by the commonly accepted QNMs of spherical resonators, which were used in the example by the Comment authors themselves.\newline

In conclusion, we find that Ref.~\cite{arXiv:1602.07278v1} adds nothing constructive to the current literature on QNMs. %, and to the use of QNMs in Purcell factor calculations. 
Using only information already contained in Ref.~\cite{1}, we have dismissed every point of criticism in the Comment. We fully stand by our results in Ref.~\cite{1} and, as far as we are aware, the results are correct.

\appendix
\section*{Appendix: Connection between the normalizations by  Muljarov \lowercase{\emph{et al.}} and Sauvan \lowercase{\emph{et al.}}}

In Ref.~\cite{1}, we presented calculations showing the connection, for general resonators, between the normalizations by Lai \emph{et al.}~\cite{Lai} and Sauvan \emph{et al.}~\cite{Sauvan} as well as the connection between the normalizations by Lai \emph{et al} and Muljarov \emph{et al.} \cite{EPL}. To complete the picture, we show here the connection between the normalizations by Sauvan \emph{et al.} and Muljarov \emph{et al.} for general resonators made from isotropic and non-magnetic materials as considered also in Ref.~\cite{1}. %As pointed out by Sauvan \emph{et al.}, with $\mathbf{J}_1=\mathbf{J}_2=0$, we have immediately an orthogonality relation for the quasinormal modes of different frequencies. 

From Eq. (S2) of the supplementary information in Ref.~\cite{Sauvan}, setting $\mathbf{j}_1=\mathbf{j}_2=0$, and adopting the notation in Ref.~\cite{1}, we have the relation
\begin{align}
0 =\; \frac{\text{i}}{2}&\int_V\mft_\mu(\mr)\cdot\big[\tlo_\mu\epsilon_\mr(\mr,\tlo_\mu) - \omega\epsilon_\mr(\mr,\omega)\big]\mft(\mr,\omega) -\frac{\mu_0}{\epsilon_0}(\tlo_\mu-\omega)\mgt_\mu(\mr)\cdot\mgt(\mr,\omega) \ud V \nonumber \\ &- \frac{1}{2\epsilon_0}\int_{\partial V} \big[\mft(\mr,\omega)\times\mgt_\mu(\mr) - \mft_\mu(\mr)\times\mgt(\mr,\omega)\big]\cdot\mathbf{n}\;\ud A,
\end{align}
where $\mft_\mu(\mr)$ and $\mgt_\mu(\mr)$ denote the electric and magnetic field QNMs, respectively, with complex (angular) resonance frequency $\tlo_\mu$. Following Muljarov \emph{et al.}~\cite{EPL,2}, the fields $\mft(\mr,\omega)$ and $\mgt(\mr,\omega)$ denote analytical continuations of the QNMs in the vicinity of $\omega=\tlo_\mu$. 
Sauvan \emph{et al.} argues that in calculations using perfectly matched layers (PMLs) the surface integral can be immediately set to zero, because it is evaluated at the (complex) coordinate transformed positions beyond the PMLs where the fields vanish~\cite{Sauvan}. Nevertheless, the Lorentz reciprocity theorem holds for all volumes $V$, and for finite sized volumes, we must keep the second term~\cite{Kathrin}. In the limit $\omega\approx\tlo_\mu$, and hence $\mft(\mr,\omega)\approx\mft_\mu(\mr)$, we follow Sauvan and co-workers and write $\omega\epsilon_\text{r}(\mr,\omega)\approx\tlo_\mu\epsilon_\text{r}(\mr,\tlo_\mu) + \eta(\mr,\tlo_\mu)(\omega-\tlo_\mu)$, where
$\eta(\mr,\omega) = \partial_\omega[\omega\epsilon_\text{r}(\mr,\omega)]$. In this limit we can then write
\begin{align}
0 =\; (\tlo_\mu-\omega)\Big[\frac{\text{1}}{2}&\int_V\eta(\mr,\tlo_\mu)\mft_\mu(\mr)\cdot\mft(\mr,\omega) -\frac{\mu_0}{\epsilon_0}\mgt_\mu(\mr)\cdot\mgt(\mr,\omega) \ud V \nonumber \\ &+\frac{\text{i}}{2\epsilon_0(\tlo_\mu-\omega)}\int_{\partial V} \big[\mft(\mr,\omega)\times\mgt_\mu(\mr) - \mft_\mu(\mr)\times\mgt(\mr,\omega)\big]\cdot\mathbf{n}\;\ud A\Big].
\end{align}
which shows that the sum of the integrals may be different from zero as $\omega\rightarrow\tlo_\mu$, in which case it becomes the norm $\langle\langle\mft_\mu|\mft_\mu\rangle\rangle$. To explore this limit, we may rewrite the expression using
\begin{align}
\mgt(\mr,\omega) = -\frac{\text{i}}{\mu_0\omega}\nabla\times\mft(\mr,\omega) 
\end{align}
and the vector Green's identity of the first kind, 
\begin{align}
\int_V (\nabla\times\mathbf{P})\cdot(\nabla\times\mathbf{Q})&- \mathbf{P}\cdot\nabla\times\nabla\times\mathbf{Q}\,\ud V =\int_{\partial V}\mathbf{n}\cdot(\mathbf{P}\times\nabla\times\mathbf{Q})\,\ud A,
\end{align}
as 
\begin{align}
\langle\langle\mft_\mu|\mft_\mu\rangle\rangle = \lim_{\omega\rightarrow\tlo_\mu}&\Big\{ \int_V\sigma(\mr,\omega)\mft_\mu(\mr)\cdot\mft(\mr,\omega)\ud V \nonumber \\
&+ \frac{\text{c}^2}{2\tlo_\mu\omega}\int_{\partial V} \Big[\mft_\mu(\mr)\times\nabla\times\mft(\mr,\omega) %\big]\cdot\ud\mA \nonumber\\
%&- \frac{\text{c}^2}{2\tlo_\mu\omega}\int_{\partial V} 
-\frac{ \omega\mft(\mr,\omega)\times\nabla\times\mft_\mu(\mr) - \tlo_\mu\mft_\mu(\mr)\times\nabla\times\mft(\mr,\omega) }{\omega-\tlo_\mu}\Big]\cdot\mathbf{n}\;\ud A\Big\},
\end{align}
where $\sigma(\mr,\omega) = \partial_\omega[\omega^2\epsilon_\text{r}(\mr,\omega)]/2\omega$. To further investigate the behavior of the last term as $\omega\rightarrow\tlo_\mu$, we follow Muljarov \emph{et al.}~\cite{EPL,2} and write
\begin{align}
\mft(\mr,\omega) &\approx %%\mft_\mu(\mr) + (\omega-\tlo_\mu)\frac{r}{\text{c}} \partial_{kr}\mft_\mu(k\mr) \\ %\frac{\partial}{\partial(kr)}\mft_\mu(k\mr)\\
%&= \mft_\mu(\mr) + \frac{\omega-\tlo_\mu}{\tlo_\mu}r\partial_r\mft_\mu(\mr)\\
%&= 
\mft_\mu(\mr) + \frac{\omega-\tlo_\mu}{\tlo_\mu}\mathbf{K}_\mu(\mr), 
\label{Eq:E_1_w_K_1}
\end{align}
in which $\mathbf{K}_\mu(\mr)=(\mr\cdot\nabla)\mft_\mu(\mr)$ as detailed in Ref.~\cite{2}. Inserting in the second integral and taking the limit $\omega\rightarrow \tlo_\mu$ (and thus $\mft(\mr,\omega)\rightarrow \mft_\mu(\mr)$), we find
%
%we find
%\begin{align}
%S = &\frac{1}{2}\left(1+\frac{\omega}{\tlo_\mu}\right)\int_V\epsilon_\text{r}(\mr)\mft_\mu(\mr)\cdot\mft(\mr,\omega)\ud V \nonumber \\
%&+ \frac{\text{c}^2}{2\tlo_\mu\omega}\int_{\partial V} \big[\mft_\mu(\mr)\times\nabla\times\mft(\mr,\omega)\big]\cdot\ud\mA \nonumber\\
%&- \frac{\text{c}^2}{2\tlo_\mu\omega}\int_{\partial V}  \big[ \mft_\mu(\mr)\times[\nabla\times\mft_\mu(\mr)] + \frac{\omega}{\tlo_\mu}\mathbf{K}_\mu(\mr) \times[\nabla\times\mft_\mu(\mr)] %\nonumber \\%&\qquad\qquad\qquad\qquad 
%- \mft_\mu(\mr)\times[\nabla\times\mathbf{K}_\mu(\mr)] \big] \cdot\ud\mA. 
%\end{align}
%Taking now the limit $\omega\rightarrow \tlo_\mu$ (and thus $\mft(\mr,\omega)\rightarrow \mft_\mu(\mr)$ in the first two integrals), we find
\begin{align}
\langle\langle\mft_\mu|\mft_\mu\rangle\rangle = &\int_V\sigma(\mr,\tlo_\mu)\mft_\mu(\mr)\cdot\mft_\mu(\mr)\ud V + \frac{\text{c}^2}{2\tlo_\mu^2}\int_{\partial V}  \big[ \mft_\mu(\mr)\times\nabla\times\mathbf{K}_\mu(\mr) - \mathbf{K}_\mu(\mr) \times\nabla\times\mft_\mu(\mr) \big] \cdot\mathbf{n}\;\ud A. 
\label{Eq:norm_w_curls}
\end{align}
Now, we can use the general result
\begin{align}
\big[\mathbf{P}\cdot\nabla\times\mathbf{Q}\big]\cdot\mathbf{n} = \mathbf{P}\cdot\partial_\mathbf{n}\mathbf{Q}-\big[\mathbf{P}\cdot\nabla\mathbf{Q}\big]\cdot\mathbf{n},
\end{align}
where $\partial_\mathbf{n}\mathbf{Q}$ denotes differentiation of each component of $\mathbf{Q}$ in the direction of the unit vector $\mathbf{n}$ and $\nabla\mathbf{Q} = \sum_{m,n} \partial_m Q_n\me_m\me_n$, to rewrite the expression as 
\begin{align}
\langle\langle\mft_\mu|\mft_\mu\rangle\rangle = \int_V\sigma(\mr,\tlo_\mu)\mft_\mu(\mr)\cdot\mft_\mu(\mr)\ud V &+ \frac{\text{c}^2}{2\tlo_\mu^2}\int_{\partial V}  \big[ \mft_\mu(\mr)\cdot\partial_\mathbf{n} \mathbf{K}_\mu(\mr) - \mathbf{K}_\mu(\mr)\cdot\partial_\mathbf{n} \mft_\mu(\mr)\big] \cdot \mathbf{n}\;\ud A \nonumber \\ 
&- \frac{\text{c}^2}{2\tlo_\mu^2}\int_{\partial V}  \big[\mft_\mu(\mr)\cdot\nabla\mathbf{K}_\mu(\mr) - \mathbf{K}_\mu(\mr)\cdot\nabla\mft_\mu(\mr) \big] \cdot \mathbf{n}\;\ud A. 
\label{Eq:norm_w_grads}
\end{align}
Last, since $\nabla\cdot\mft_\mu(\mr)=\nabla\cdot\mathbf{K}_\mu(\mr)=0$ on the calculation domain boundary, we can use the general product rule
\begin{align}
\nabla\times\big[\mathbf{P}\times\mathbf{Q}\big] = \mathbf{P}\cdot\nabla\mathbf{Q}-\mathbf{Q}\cdot\nabla\mathbf{P}+\mathbf{P}\big[\nabla\cdot\mathbf{Q}\big]-\mathbf{Q}\big[\nabla\cdot\mathbf{P}\big],
\end{align}
to rewrite the third integral as the flux of the curl through a closed surface, which vanishes by virtue of Stokes' integral theorem. The normalization then takes the exact form presented by Muljarov \emph{et al.}~\cite{EPL}.

%%%%%%%%%%%%%%%%%%%%%%% References %%%%%%%%%%%%%%%%%%%%%%%%%

\end{document}